# Electrically Controlled Interfacial Charge Transfer Induced Excitons in MoSe$_2$-WSe$_2$ Lateral Heterostructure

Baisali Kundu[1], Priyanka Mondal[1,2], David Tebbe[2], Md. Nur Hassan[3], Suman Kumar Chakraborty[1], Marvin Metzelaars[4], Paul Kögerler[4], Debjani Karmakar[3,5,6], Christoph Stampfer[2], Bernd Beschoten[2], Lutz Waldecker[2]*, Prasana Kumar Sahoo[1]**

[1]*Materials Science Centre, Indian Institute of Technology, Kharagpur 721302, India*
[2]*2nd Institute of Physics and JARA-FIT, RWTH Aachen University, Aachen 52056, Germany*
[3]*Department of Physics and Astronomy, Uppsala University, Uppsala 75120, Sweden*
[4]*Institute of Inorganic Chemistry, RWTH Aachen University, 52074 Aachen, Germany*
[5]*Technical Physics Division, Bhabha Atomic Research Centre, Mumbai 400085, India*
[6] *Homi Bhabha National Institute, Trombay, Mumbai 400094, India*

*waldecker@physik.rwth-aachen.de
**prasana@matsc.iitkgp.ac.in

## Abstract

Controlling excitons and their transport in two-dimensional (2D) transition metal dichalcogenides (TMDs) heterostructures is central to advancing photonics and electronics on-chip integration. We investigate the controlled generation and manipulation of excitons and their complexes in monolayer (1L) MoSe$_2$-WSe$_2$ lateral heterostructure (LHS), directly grown via water-assisted chemical vapor deposition. Using a field-effect transistor design by incorporating a few-layer graphene back gate, single-layer graphene edge contact and encapsulation with few-layer hexagonal boron nitride, we achieve precise electrical tuning of exciton complexes and their transfer across 1D interfaces. At cryogenic temperatures (4 K), photoluminescence and photocurrent maps reveal the synergistic effect of local electric field and interface phenomena in the modulation of excitons, trions, and free carriers. We observe spatial variations in exciton and trion densities driven by exciton-trion conversion under electrical manipulation. The first-principle density functional theory calculation reveals significant band modification at the lateral interfaces and graphene-TMDs contact region. Furthermore, we demonstrate the versatility of 2D TMDS LHS in hosting and manipulating quantum emitters, achieving precise control over narrow-band emissions through modulating carrier injection and electrical biasing. This work extends the boundary of the present understanding of excitonic behaviour within lateral heterojunctions, highlighting the potential for controlled exciton manipulation across 1D interfaces and paving the way for next-generation electro-optical quantum devices.

## Introduction

Manipulation of exciton transport in two-dimensional (2D) heterostructure is essential for revolutionizing quantum communications, computing, and sensing applications with unique functionality, precision, and speed.[1–5] Especially, exciton propagation at room temperature with significant diffusion length is the need of the hour. Heterostructures composed of monolayer (1L) transition metal dichalcogenides (TMDs) can provide promising avenues for manipulating the formation and functionalities of long-lived excitons and trions.[1,6–9] An external electric field can modify the electronic profile of 2D heterostructures hosting different excitons, thereby passively influencing exciton characteristics such as absorption spectrum, spatial distribution, and lifetime.[10–12] This electrostatic tunability enables the control over charge-carrier density within the material, facilitating neutral-to-charged exciton conversion and affecting carrier and valley dynamics, including recombination and transport.

Charged excitons (trions) can also be controlled directly through electrical probing, possessing a longer lifetime that provides significant time for electrical manipulation before undergoing spontaneous radiative recombination.[13,14] In this context, several 2D vertical heterostructures (VHS) have been exploited in controlling interlayer excitons by the artificial engineering of the crystal lattice,[15] strain[16,17], dielectric[17,18], and valley degree of freedom[4,19]. Typically, the out-of-plane dipolar interface exciton in VHS shows anomalous diffusion due to dipolar repulsion. In contrast, a unidirectional exciton flow can be achieved in lateral geometry depending on the potential gradient between two adjacent materials.[4,20–22] However, the exact origin of different exciton species and their transport requires tedious research and challenges in precisely engineering the interfaces. Conventional transfer processes often introduce unwanted contamination, hence the reliability of the measurement process and lack of scalability.[23,24]

In this context, 2D lateral heterostructures (LHS), where spatially separated heterogeneous materials are stitched epitaxially in one plane, can offer a unique platform for directly mapping the exact origin and characteristics of charged carriers, excitons, trions, and energy transfer across the 1D interface. Especially, LHS with multiple junctions within a single layer enables the spatial design of complex devices.[25,26] The exploration of the interfacial properties can be accomplished through various experimental techniques[21,27,28]. Understanding the 1D interface-induced exciton physics is essential for the design of directional excitonic transistors utilizing 2D TMDs LHS. However, the transfer characteristics, reliability, and performance of excitonic device configurations can be significantly influenced by the quality of 1D interfaces in LHSs. Moreover, unlike VHS, LHS with pristine interfaces can only be fabricated via direct growth.

In this context, chemical vapor deposition (CVD) is a simple and robust method for directly fabricating different 2D LHSs, with precise control over critical parameters such as crystallinity, domain size, and electronic quality, ensuring a high yield.[25,29] Encapsulation of 2D TMDs within a few layers of hexagonal boron nitride (hBN) significantly improves the photoluminescence (PL) emission characteristics as it lessens the impact of environmental perturbations, reduces the inhomogeneous PL broadening, and facilitates resolving optical transition energy.[30–33] Incorporating graphene as contact electrodes to TMD promotes efficient charge carrier transfer processes.[34,35] Instead of a thicker $SiO_2$ layer, a graphite back gate to the TMDs offers several advantages, including enhancing carrier mobility, low trap density surface, improvement of electrical stability, heat dissipation due to high thermal conductivity, etc.

In this study, we fabricated 2D FET consisting of an hBN-encapsulated CVD-grown 1L $MoSe_2$-$WSe_2$ LHS as the channel, graphene as the contact electrode and few-layer graphene (FLG) as the back gate. A top-down dry-transfer process achieves the multiple-stacked device structure with ultimate precision. We conduct a detailed investigation of electrically controlled PL and photocurrent from the 2D $MoSe_2$-$WSe_2$ LHS-based FET device at cryogenic temperatures. In the FET geometry, we observe high-density quantum emissions that can be further tuned via external electric fields, enabling a detailed assessment of excitonic features and more precise observation of excitonic and localized state emissions.

**Result and Discussion**

A one-pot water-assisted CVD strategy is used to synthesize the MoSe$_2$-WSe$_2$ LHS. Single-layer LHS flakes were dry-transferred using a PC/PDMS stack from the as-grown SiO$_2$/Si$^{++}$ substrate on hBN flakes of thickness 30 nm.[36] A few layer graphene flake at the back of the bottom hBN serve as a back-gate electrode for the entire device (Figure 1a). The details of the transfer process are explained in the sample preparation and device fabrication section. Single-layer graphene flakes are placed on MoSe$_2$ and WSe$_2$ domains to serve as contact electrodes for the individual domains. An insulating small hBN flake has been used beneath the graphene electrode for MoSe$_2$ domain, as depicted in Figure 1b. Further, electrodes were developed to contact the graphene flakes using electron beam lithography followed by deposition of Ti (5 nm)/Au (90 nm) via e-beam evaporation. Figure 1c illustrates the differential reflection constant variation at different positions of the MoSe$_2$-WSe$_2$ LHS flake at room temperature. Exciton resonance positions are determined to be 759 nm for WSe$_2$ and 796 nm for MoSe$_2$, respectively. The reflection constant is calculated in the reflection measurement under the white light by subtracting the background reflection spectra as $(R_{sample} - R_{SiO_2/Si})/R_{SiO_2/Si}$. The distinctive asymmetric Fano-like line shape is highly noticeable in MoSe$_2$, indicating interference between a broad background and sharp excitonic resonances, whereas it appears nearly symmetric in WSe$_2$.[37,38]

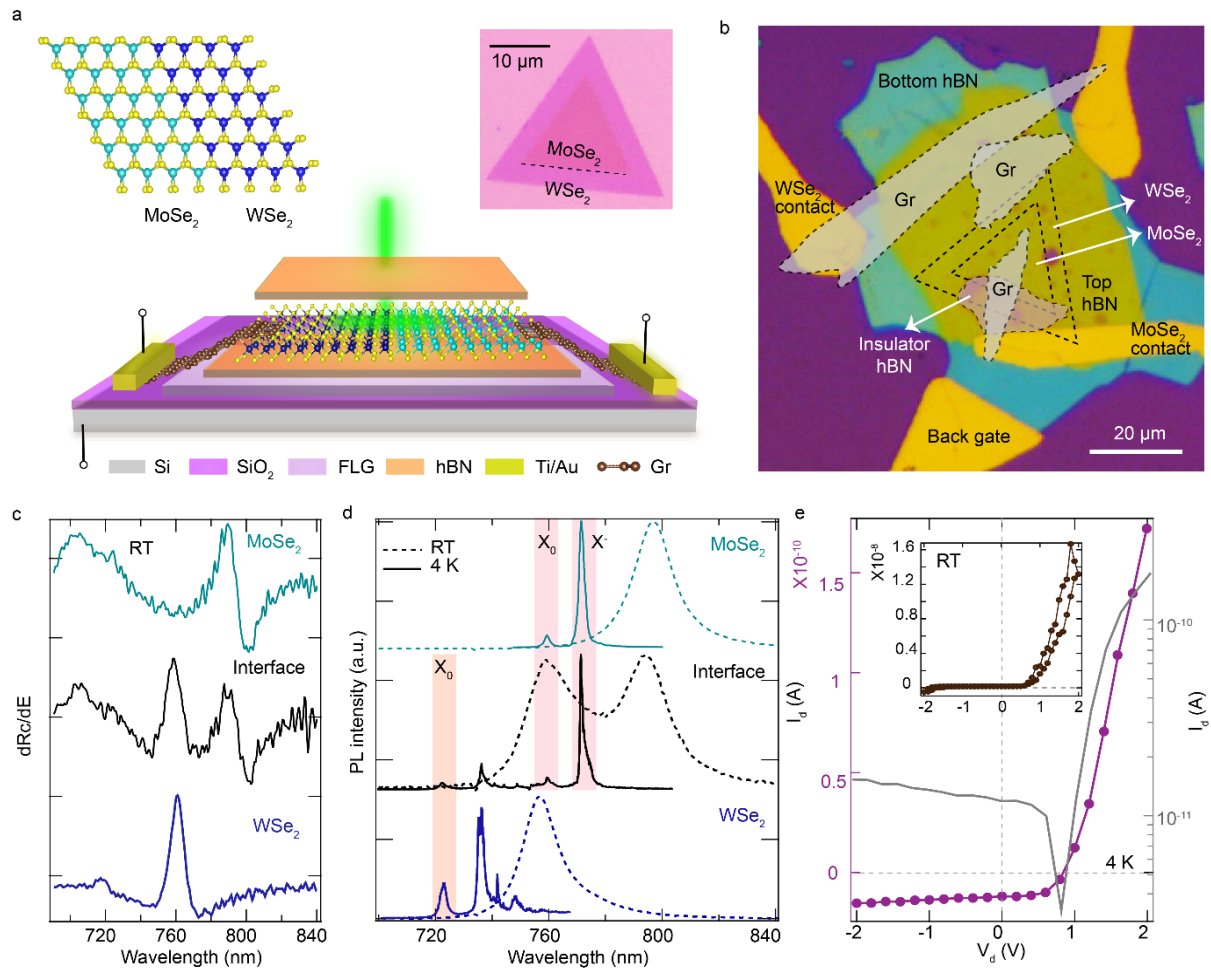

**Figure 1. a.** An atomic-ball model of a monolayer (1L) MoSe$_2$-WSe$_2$ lateral heterostructure (LHS) and a schematic representation of the optoelectrical device based on an hBN-encapsulated one-junction MoSe$_2$-WSe$_2$ LHS; the graphene layers are placed for the contacting of the individual TMD domains and the bottom few-layer graphene (FLG) serves as a back-gate; An optical image of the flake is shown top right. **b.** An optical micrograph showing the device configuration; MoSe$_2$-WSe$_2$ flake, graphene, and hBN regions are outlined and shaded for clarity. **c.** Room temperature differential reflectivity

measurements showing a differential reflection constant for WSe$_2$, MoSe$_2$, and the interface region. **d.** PL spectra at 4K (solid line) and room temperature (dotted line) show the emission characteristics of the individual TMDs layer and at the interface. **e.** A typical diode characteristic without gate bias confirms the device as an active p-n junction at 4 K (and RT in inset) under dark conditions.

The room temperature PL and Raman spectra with 532 nm laser excitation (spot size 0.4 µm)[39] are shown in Figure 1d and S1. The PL resonant wavelengths at 756 nm and 796 nm depict direct A-excitonic transition from the pristine WSe$_2$ and MoSe$_2$, respectively. It reveals distinct PL features of the two domains with line widths of approximately 15 nm (32 meV) for WSe$_2$ and 14 nm (28 meV) for MoSe$_2$. Additionally, superimposed PL features of both domains are observed at the interface. For 1L MoSe$_2$, illumination favors the direct bright exciton ($X_0$) transition as the dark exciton states at $K$ and $K'$ point of the Brillouin zone positioned energetically higher than the bright states in contrast to WSe$_2$ at room temperature. As a result, the relative PL intensity is higher for MoSe$_2$ than for WSe$_2$[40–42]. However, due to reduced thermal effect, excitonic features and the defect or trap state-related emissions can be well resolved at cryogenic temperature. At 4K, the PL spectra of 1L WSe$_2$ and MoSe$_2$ reveal the $X_0$ at 723 nm and 759 nm (Figure 1d), with line widths of 2.5 nm (5 meV) and 1.8 nm (4 meV), respectively, comparable with the exfoliated 1L TMDs[43,44]. This narrow PL line width is attributed to the high crystalline quality of the as-synthesized 2D TMDS encapsulated with hBN layers, effectively shielding the flake from environmental perturbations and minimizing inhomogeneous broadening. MoSe$_2$ shows a strong $X^-$ intensity with a line width of 2.1 nm (3.7 meV) lying 12.3 nm (26 meV) below $X_0$[43,45]. On the other hand, WSe$_2$ exhibits a series of sharp excitonic transitions between 735 to 764 nm. Localized emitters in WSe$_2$ can be seen as a result of intrinsic defects, traps or strains, whereby the exact origin of such emitters remains to be investigated.[46–50] Furthermore, we evaluated the transport characteristics (current/bias voltage traces) of this device under dark conditions, revealing a unidirectional current flow indicative of an active p-n diode at both room temperature as well as 4K (Figure 1e and S2). The maximum channel current of 165 pA was extracted at $V_d$=2 V at 4 K, showing a two-fold reduction compared to the room temperature measurement. The reduced current at 4K can be understood from the thermal-activated transport equation, $J \propto AT^2 \exp(-\frac{qV_b}{kT})$]; where J is the current density which decreases at lower temperature ($T$), A is the channel area and $V_b$ is the barrier height. As the temperature decreases, the probability of electron occupancy above the Fermi energy level decreases. Consequently, the number of free carriers in the conduction band is lower compared to room temperature, resulting in a reduced flow of current.

The 2D-LHS based FET with a thin graphite gate introduces additional charge carriers that can modulate the electrostatic doping within channel materials. The contour plots of typical PL intensities, dependent on the external bias $V_g$, demonstrate the presence of various exciton species for WSe$_2$ and MoSe$_2$ at 4K, as depicted in Figures 2a and 2c (in addition with Figure S3). At $V_g$ = -0.5 V, the WSe$_2$ domain is almost charge neutral[47], whereas MoSe$_2$ shows a prominent n-type feature with $X^-$ in addition to $X_0$. The WSe$_2$ domain shows the characteristic of p-type doping at -2.5 V and n-type doping at 1 V while the $X_0$ is quenched at this $V_g$ values due to the conversion of excitons into trions. Additionally, intervalley and intravalley trions ($X_1^-$, $X_2^-$) at 736 nm and 741 nm appear in the WSe$_2$ domain at the high positive $V_g$ suppressing $X_0$ peak intensity[47]. The narrow line emissions at longer wavelength than the trion resonance are further characterized and discussed in a separate section. The spatial PL profile of the WSe$_2$ domain, spectrally integrated around $X_0$, is depicted in Figure 2b at two voltage extremes (1 V and -2.5 V) and the charge-neutral $V_g$ of -0.5 V. The white dashed lines mark the junction region with the left-hand side as WSe$_2$. The spectral profiles are found to be mostly similar whereas, intensity is found to have slight variation due to the relative difference in carrier concentration for different $V_g$.

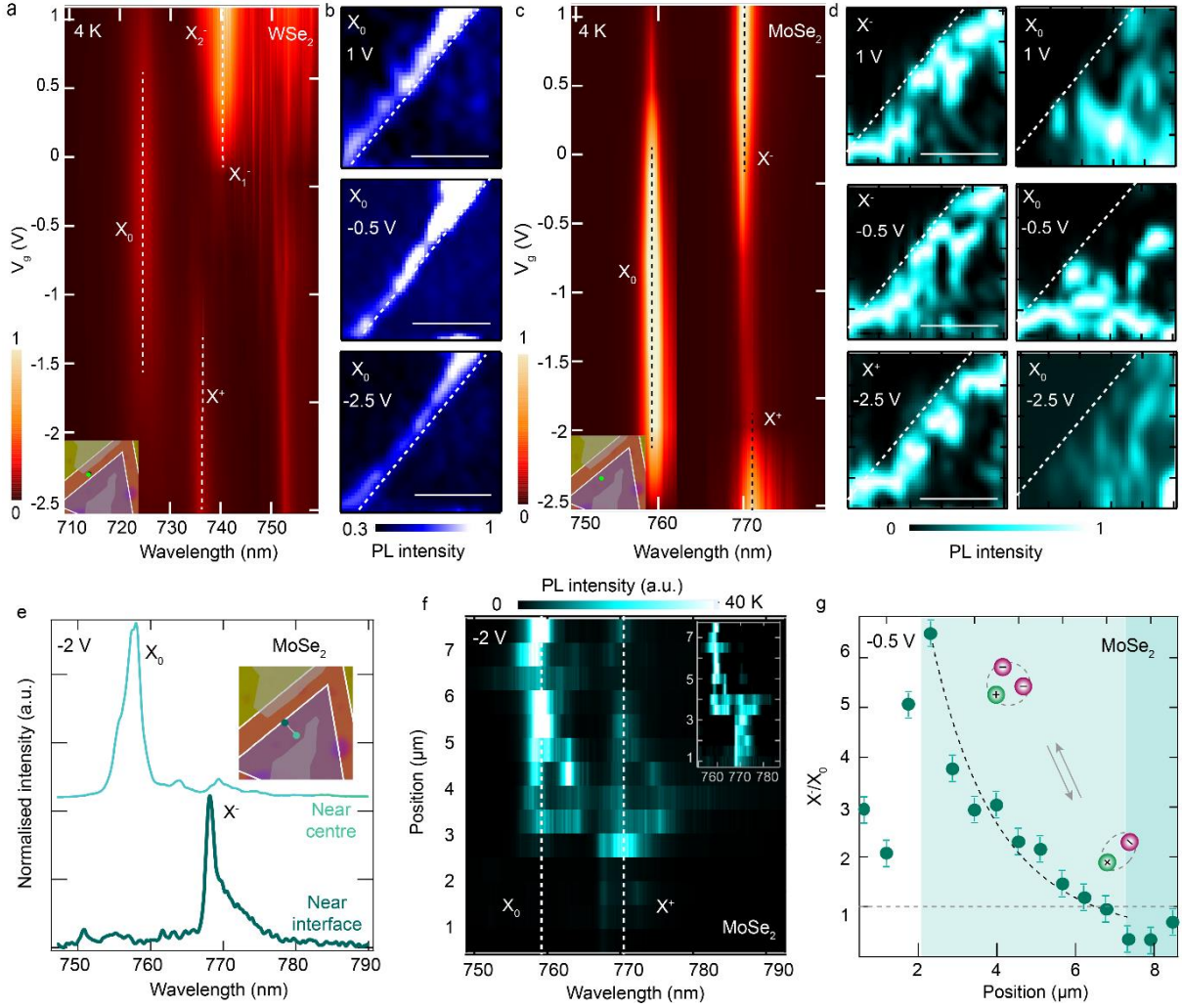

**Figure 2 a.** Gate bias-dependent PL intensity profile for 1L WSe$_2$; inset shows the corresponding optical image. **b.** Spatial PL mapping indicating the WSe$_2$ region at the $X_0$ wavelength at different $V_g$; the white dotted line follows the interface shown in the inset of a. **c.** $V_g$-dependent PL intensity profile at MoSe$_2$ showing distinct $X_0$, $X^\pm$ peaks . **d.** PL profile at MoSe$_2$ showing the spatial doping difference at $X_0$ and $X^\pm$ while $X_0$ predominating towards the center and $X^\pm$ towards the interface. **e.** PL spectra near the interface and the center region, as indicated in the inset, reveal the dominance of $X_0$ or $X^\pm$ emissions at -2 V of $V_g$. **f.** A line scan denoting the spatial region where the exciton emission intensities are high at $V_g$= -2 V; normalised line intensity profile is shown in the inset. **g.** The intensity ratio of $X_-$ and $X_0$ shows a decremental trend towards the center (scan position 8 µm), followed by an increase (from 8 µm onwards) towards the subsequent opposite boundary of MoSe$_2$; the dotted line follows an exponential fit. Scale bars in a, c insets are 10 µm and b, d are 5 µm.

The normalized PL spectral map from the MoSe$_2$ domain shows the presence of well-defined $X_0$, $X^+$ and $X^-$ bands at different $V_g$ (Figure 2d and S4). An intriguing interface phenomenon was observed for the charged exciton. At the lateral interface, the intensities of $X^+$ and $X^-$ are strong and diminish towards the center of MoSe$_2$, whereas, $X_0$ follows the opposite trend. This phenomenon is seen for all gate voltages. However, at $V_g$ =-0.5 V the distribution of the $X_0$ is observed almost all over the MoSe$_2$ domain because of the strong $X_0$ intensity as compared to the charged one. Interestingly, it can be seen that the PL spectrum in the proximity of the interface within MoSe$_2$ is dominated by the formation of trion. (Figure 2e). The line scan across the junction in a contour plot (Figure 2f) clearly illustrates a gradual transition from $X_0$ towards $X^+$ at $V_g$ = -2 V while approaching the interface, as indicated by the dotted line shown in Figure 2e. A sharp-intensity contrast from $X_0$ to $X^+$ indicating a shift in the spectral distribution of trions at the proximity of the interface. The PL peak intensity ratio of $X^-$ and $X_0$ decreases at -0.5 V (Figure 2g). Here, the contribution related to an increment of trion and spatial decrement of exciton concentration towards the interface can be explained via a spatial model of free-charge carriers across a p-n junction.[51]

$$dn = n_0 e^{-x/L}$$

where, $n_0$ is the equilibrium density of majority carriers in WSe$_2$, which diffuse over the length of $x$ through the adjacent MoSe$_2$ and $L$ is the maximum diffusion length. If the transferred carriers are responsible for the increment of the trion density near the junction, the modified trion density towards the junction can be expressed as,

$$n_T = n_T^0 + n_0 e^{-x/L}$$

and the exciton density as $\quad n_X = n_X^0 - n_0 e^{-x/L}$.

By applying the equilibrium condition $n_T + n_X = n_T^0 + n_X^0$, the density ratio becomes

$$n_T/n_X = n_T^0 \left(1 + ae^{-\frac{x}{L}}\right)/n_X^0 \left(1 - be^{-\frac{x}{L}}\right); \quad a = \frac{n_0}{n_T^0}, b = \frac{n_0}{n_X^0}$$

At x<<L near the junction, $\quad n_T/n_X \approx 1 + ce^{-x/L} + de^{-2x/L}; \quad c = n_0(\frac{1}{n_T^0} + \frac{1}{n_X^0}), d = \frac{n_0}{n_T^0 n_X^0}$

The final equation obtained from the microscopic model shows a strong alignment (dashed line) with the experimental result shown in Figure 2g. The arbitrary points near the interface may be attributed to the combined influence of mobile carriers receding, resulting in the formation of space charge region and carrier depletion due to diffusion.

To probe the interface effect in the formation of excitons, low-temperature scanning photocurrent mapping was performed using a 532 nm laser of 0.5 µW (Figure 3a, b). Although a uniform laser illumination was maintained throughout the scanning process, the WSe$_2$ region yields a higher photocurrent than the MoSe$_2$ region, suggesting a significant abundance of free carriers even at 4K. It can be seen that the high photocurrent zone is aligned with the graphene edge placed on the WSe$_2$ domain, as depicted in Figure 3a. A line scan across the interface clearly distinguishes the photocurrent profile across the junction (Figure 3c). Contrary to the expected high photocurrent across the p-n junction, the observed high photocurrent at WSe$_2$ may be attributed to the synergistic effect of high PL yield and interfacial charge transfer process. In general, WSe$_2$-MoSe$_2$ possesses a type-II band alignment, and the Fermi energy ($E_F$) aligns at a steady state condition through majority carrier diffusion. The junction potential can be altered by applying an external electric field, whereas forward biasing decreases the depletion region, easing the carrier flow. Under the positive $V_g$, $E_F$ shifts towards the conduction band edge ($E_C$), resulting in an increasing free electron concentration in the MoSe$_2$ domain, whereas negative $V_g$ electrostatically hole dope the WSe$_2$ domain. The increased photocurrent generation in WSe$_2$, despite a relatively higher bandgap compared to MoSe$_2$, can be attributed to the synergistic effect of the efficient charge transfer process to graphene electrode and heavy density of localised defect states as seen in the PL spectra (Figure 2a). However, encapsulation of MoSe$_2$ predominantly eliminates trapping carriers. These activated defect or trap states in WSe$_2$, located within the forbidden energy gap, capture the minority carriers under non-equilibrium conditions (Figure 3d).[52,53] Upon light irradiation, the photogenerated carriers (electron or hole) are involved in recombination with trap charges.

Meanwhile, the remaining opposite charge carriers give rise to the current density, especially in the WSe$_2$ domain.[54,55] The trap states, upon illumination, collectively participate in the radiative carrier recombination, as evidenced by PL and the photocurrent map facilitated by the remaining free carriers. The free carriers near the interface can transfer to the adjacent TMDs domain due to band bending with an appropriate applied bias. On the other hand, the excitons in MoSe$_2$ capture the excess free carriers, forming trions. Depending on the gate bias, WSe$_2$ produces free electrons or free holes to travel across the junction. Consequently, the charge state of the trion in MoSe$_2$ is determined by the charge of the diffused carrier (electron or hole). Thus, with the positive $V_g$, the trion position on MoSe$_2$ near the interface exhibits a shift in wavelength compared to the negative $V_g$, indicating the presence of two distinct $X^+$ and $X^-$ (Figure 2c).

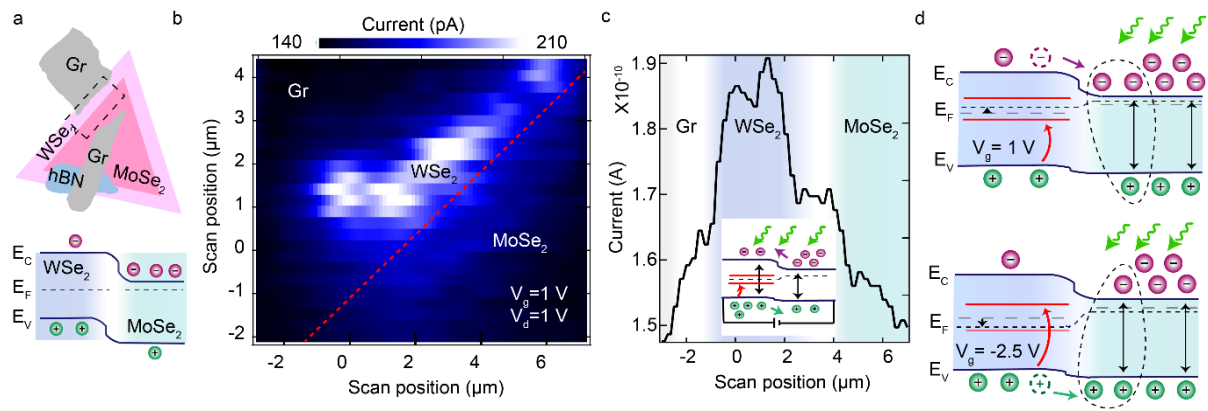

**Figure 3. a.** A diagram of the lateral heterostructure flake featuring graphene and hBN contacts emphasizes the area of scanning for photocurrent measurement; A band diagram of the p-n junction at the equilibrium condition (bottom). **b.** A photocurrent map over the dotted region under a forward bias of 1 V at positive $V_g$; the red dotted line represents the 1D lateral interface. **c.** A line scan of the photocurrent across the junction under the same bias condition; The inset band diagrams indicate the direction of the charge flow at that condition. **d.** Proposed band diagrams illustrate the formation of distinct charged excitons near the interface at two different polarities $V_g$ under forward bias; the red solid line signifies the trap energy levels capturing the minority carriers (holes for positive $V_g$ and electrons for negative $V_g$).

To obtain a comparative picture of the electronic states of the lateral MoSe$_2$-WSe$_2$ heterostructure and its configurations with graphene, we have conducted calculations of the electronic band structure of five systems, *viz*., (a) 1L MoSe$_2$, (b) 1L WSe$_2$, (c) 1L MoSe$_2$ with graphene top contact, (d) 1L WSe$_2$ with graphene top contact, (e) 1L MoSe$_2$-WSe$_2$ LHS. Figure 4 a and b present the electronic band dispersions of 1L MoSe$_2$ and WSe$_2$, projecting a direct bandgap of ~ 1.48 and 1.59 eV, respectively, which matches earlier reports.[25] The corresponding experimental analogue of MoSe$_2$ and WSe$_2$ with top graphene contact is depicted in Figure 4c and d, where the linearly dispersive graphene levels form mid-gap states. Moreover, the shifts of the $E_F$ with respect to both the pristine TMDs and graphene as top contact is towards the conduction band (Table S1). The energetic placement of VBM (-0.04 eV) and CBM (1.18 eV) of MoSe$_2$/graphene are staggered with respect to the VBM (-0.34 eV) and CBM (1.02 eV) of WSe$_2$/graphene. It keeps the trend of more n-type doping for MoSe$_2$ intact. Therefore, the possibility of negatively charged trions will be more towards the MoSe$_2$ side of the LHS, as also seen in the experiment. The atom-projected band structures of the MoSe$_2$-WSe$_2$ lateral interface (Figure 4e), constituted of the 4×4×1 1L of the pristine systems, reveal that the levels at the VBM and CBM are populated mainly by the W-5$d$ and Mo-4$d$ states, respectively, with the direct band gap shifting from K to within Γ-K. A type-II band-alignment was also predicted at the lateral interface.

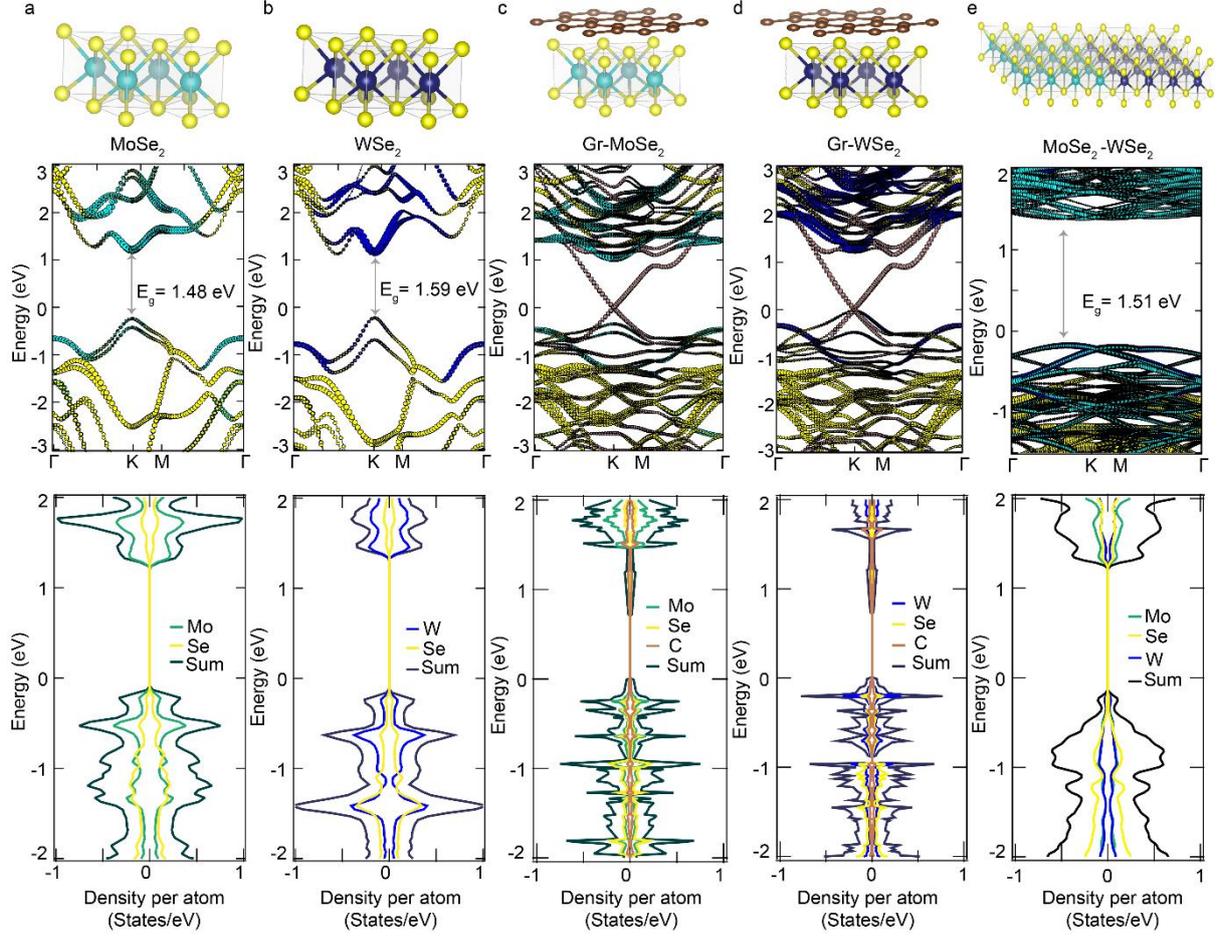

**Figure 4.** Atomic projected band structure by spin-polarized DFT and partial density of states (PDOS) focusing on the outermost orbitals of Mo 4d, W 5d, Se 4p and C 2s within **a.** 1L MoSe$_2$, **b.** 1L WSe$_2$ **c.** graphene-MoSe$_2$ **d.** graphene-WSe$_2$ and **e.** 1L-MoSe$_2$-WSe$_2$ LHS interface.

We now turn to the narrow emission lines which have been observed in the PL spectra in the WSe$_2$ domain. These have been characterized via different biasing and non-biasing conditions in conjunction with various excitation powers. We observed position-dependent sharp emitters within the WSe$_2$ domain (Figure 5) with line widths ranging from 400 to 900 µeV.[50] It is essential to tune the quantum emissions spectrally for applications pertaining to quantum information. We have observed gate tunable quantum emission features in the 2D-LHS FET configuration associated with intrinsic defect states in WSe$_2$ (Figure 5a, b and S5). It may have originated due to the variation in the band alignment of graphene to WSe$_2$ (Figure 4b, d) that changes the carrier concentration and charge states of these quantum emitters. At negative $V_g$, holes from the WSe$_2$ transfer to graphene, while at positive $V_g$ electrons can transfer from graphene to WSe$_2$. Therefore, the hole concentration is reduced at a negative $V_g$ while there exists an electron surplus at positive $V_g$.

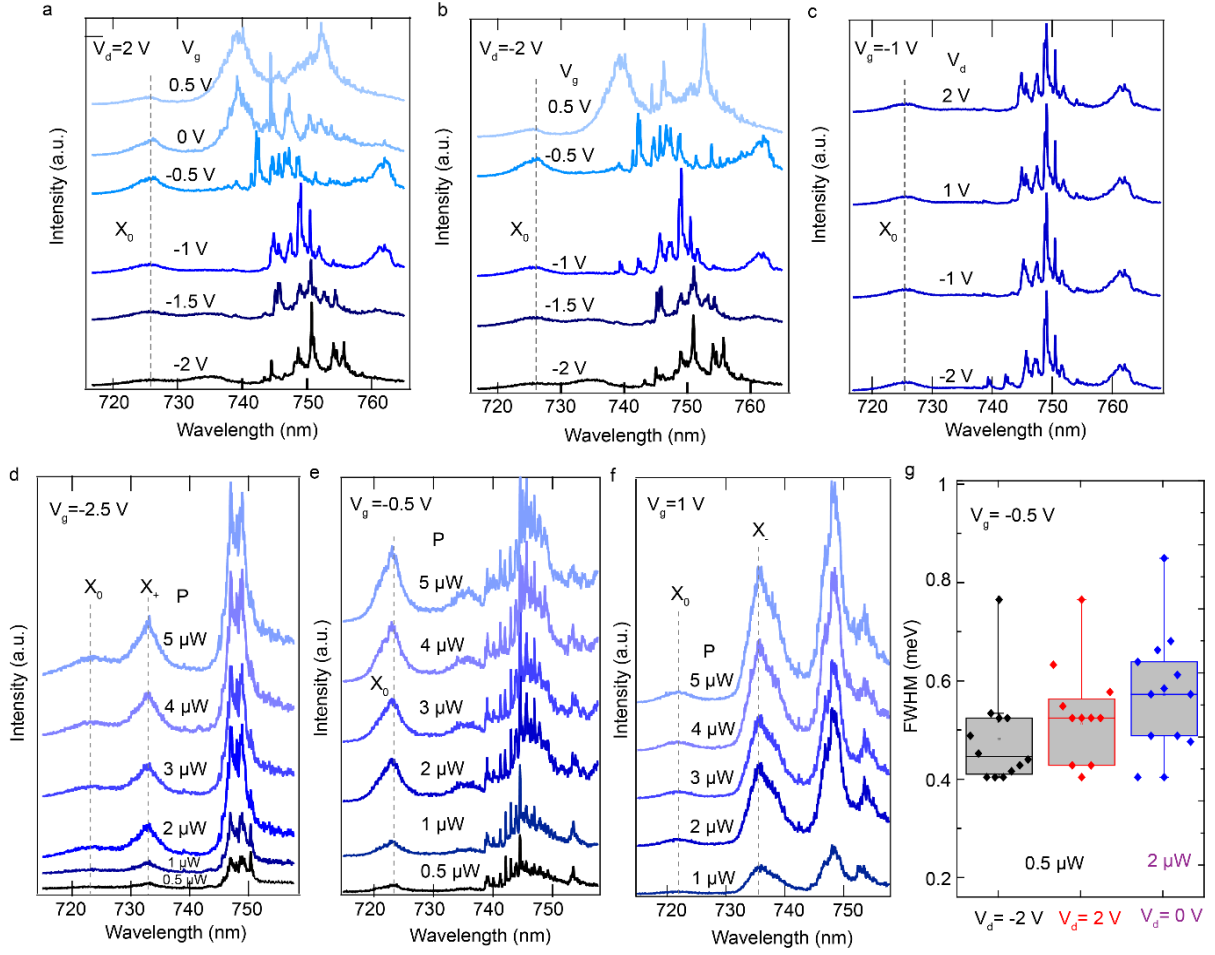

**Figure 5 a-b.** Tubnable localized quantum emitter characterization: Low-temperature PL emission profile from the 2D-LHS FET device for different $V_g$, at constant $V_d$, demonstrating the tunability of the localized emitters. **c.** $V_d$-dependent PL emission at particular $V_g$ showing the stability of these quantum emitters for different applied lateral electric fields and **d-f.** Sharp emission characteristics on incident laser power at three different $V_g$, showing the saturation behavous of such quantum emitters. **g.** A statistical analysis from different quantum emitters' line width profiles measured at different parameters while $V_g$ set at -0.5 V.

With positive applied $V_g$, the sharp emission peaks are merged and give rise to broader emission characteristics. On the contrary, under negative $V_g$, the holes reduce the screening and hence facilitate strict localization of $X_0$ in the WSe$_2$ domain, leading to the occurrence of multiple excitonic resonances[56]. We also attempted to identify the sharp emissions by adjusting the bias voltage. Interestingly, no significant change was observed in the peak positions with varying $V_d$ [Figure 5c, S6], which could be attributed to the fact that band offset does not influence trap exciton binding energies unless it involves the interface exciton[57,58]. Moreover, the emissions display a linear response to excitation power under a fixed bias within the power range used for the experiment (Figure 5d-f, S7), as reported earlier[50]. We observed variations of these localized quantum emissions alongside varying intensities, line widths, and peak shifts. A statistical analysis is presented in Figure 5g related to the change in the line widths of localized emitters. The large variation of the line widths under varying $V_d$ and laser power corroborated the synergistic effects of charge transfer and the different sizes of the electronic wave-function from the localized states. Thus, the electrically tuned PL characteristic of the narrow line emissions from localized states in WSe$_2$ in FET geometry, suggests that CVD-grown MoSe$_2$-WSe$_2$ LHS is an enticing platform to design high density and position-based programmable sharp emitters. Further investigations in this direction are expected to enhance our understanding of the exact nature of these localized states and the performance metrics of such quantum emitters, paving the way for advancements towards realizing programmable single-photon emission for future quantum communication.

**Summary**

The study highlights the existence of different excitonic complexes and their controlled generation via electrical doping in CVD-grown 1L MoSe$_2$-WSe$_2$ LHS. An optimised pick-up process and subsequent transfer of multi-material layers onto FLG facilitates complex device fabrication, enabling the precise tuning of the optoelectronic features. Unlike vdW heterostructures, the spatial variations in doping density within the 2D lateral heterojunction provided the opportunity to directly probe the intriguing carrier dynamics, free charge transfer across the interface, and the formation of interfacial exciton complexes in the LHS. Low-temperature photocurrent measurement and first-principle DFT calculation were further employed to uncover the underlying phenomenon associated with photo-excited and free-charge carriers. In addition, we demonstrated the possibility of realizing programmable quantum emitters by suitably engineering the 2D FET device design where external electric fields can be used to modulate the carrier transfer dynamics. Such high-density electrically tunable quantum emission is promising for further exploration towards their use as single photon sources. These findings not only provide a comprehensive understanding of engineering exciton dynamics within atomically stitched lateral heterointerfaces but also pave the path for future design of 2D devices for optoelectronic, photonic and quantum technology.

**Materials and methods**

**Synthesis of MoSe$_2$-WSe$_2$ LHS**

To synthesize single-junction 1L MoSe$_2$-WSe$_2$ LHS, the one-step water-assisted CVD process was employed with the mixture of MoSe$_2$ and WSe$_2$ bulk precursors.[25] The selective growth of individual TMD layers was realized simply via changing carrier gases at a relatively high temperature (~ 1050 °C), and the growth substrates were kept in the temperature range of ~800 to 850 °C. In the initial growth stage, the presence of wet-N$_2$ gas facilitated the oxidation of solid precursors, selectively promoting Mo-related precursor evaporation and leading to the formation of the MoSe$_2$ domain. Switching to reducing gas such as H$_2$ (5%) in Ar terminates, the W-based sub-oxides can evaporate and contribute to the formation of WSe$_2$ at the same temperature. The spatial width, shape, size, and degree of alloying within individual domains are controlled independently.

**Transfer process and device fabrication**

A dry transfer approach was employed to stack multiple layers for the device fabrication. A stack of polypropylene carbonate (PC) on polydimethylsiloxane (PDMS) was mounted on a clean glass slide and heated up to 160 ºC for 5 min. This elevated temperature allowed the PC to align well with the PDMS, resulting in strong adhesion between them. Under a long working distance objective and at a temperature of 110°C, the stack is sequentially placed on exfoliated hBN. The whole stack was picked up occurring at 110°C. Subsequently, the complete stack was placed onto the targeted CVD flake grown on SiO$_2$. Upon heating to 70 °C, hBN can detach and pick up the flake from SiO$_2$. A similar strategy was applied to pick up the single-layer graphene for the contact electrode and top hBN as a protective layer. The entire stacking process was carried out in a glovebox with a nitrogen atmosphere to avoid oxidation and moisture contamination. After picking up the desired layers, the entire stack was placed on a clean SiO$_2$ substrate and heated to 180 °C. At this temperature, the PC turned to be flexible, and the stack-PC assembly detached from PDMS while the transfer arm lifted against the stack-contained substrate. The remaining PC layer was removed from the desired material stack by treating it in chloroform for 15 min. The cleaned stack was further annealed at 180 °C for 5 minutes to avoid the impact of any chemical residue. Electron beam lithography was utilized to fabricate the mask for the electrode contacts to the sample. Subsequently, a thin layer of titanium (5 nm) followed by a thicker layer of gold (90 nm) was deposited using e-beam evaporation.

**Experimental setup and procedure for PL and device characterization**

All experiments were performed in a cryostat (AttoLiquid), equipped with XY scanners. Samples were cooled down to 4 K for PL and transport measurements. The optical setup includes a Tungsten halogen white light source for the reflection measurement and a 532 nm continuous wave (CW) laser source for PL and photocurrent measurements.[39] An objective lens of N.A. 0.8 (100X) has been used to tightly focus the light beam and collect the signal in back reflection geometry. After passing through the spectrometer (with a 1200l/mm grating), the signal was mapped to the 2D array of a CCD. The spectral resolution of the setup was 300 µeV. Photocurrent measurements were done via a 2D scan with a step size of 0.25 µm along the x- and y-axes. The electrical characterization was performed using a Keithley 2611A SourceMeter.

**DFT Calculations**

The spin-polarized plane-wave calculations were carried out with the help of projector-augmented wave (PAW) pseudopotentials with spin-orbit coupling (SOC). We have used generalized gradient approximation (GGA) with Perdew-Burke-Ernzerhof (PBE) exchange-correlation functional, as implemented in the Vienna ab initio simulation program (VASP).[59] Van-der-Waals (VdW) interactions are taken into account in terms of the semiempirical dispersion potential, representing the dipolar interactions. It is included within the DFT energy function using the DFT-D3 method of Grimme.[60] A Monkhorst-Pack k-grid of dimension 5×5×3 is used for sampling the Brillouin zone, and the cut-off for the plane wave expansion is kept at 500 eV. The energy convergence for all self-consistent field calculations is kept as $10^{-5}$ eV, and the structural optimizations were performed using the conjugate gradient algorithm until the Hellmann-Feynman forces on each ion were less than 0.01 eV/Å.

**Acknowledgement**

P.K. acknowledge the Department of Science and Technology (DST), India (Project Code: DST/NM/TUE/QM-1/2019; DST/TDT/AMT/2021/003 (G)&(C)), and ISIRD start-up grant (ISIRD/2019-2020/23) from the Indian Institute of Technology Kharagpur. M.N.H. and D.K. would like to acknowledge the Bhabha Atomic Research Centre (BARC), Mumbai, ANUPAM supercomputing facility for computational resources. D.T., C.S., B.B., and L.W. acknowledge the support by the Deutsche Forschungsgemeinschaft (DFG, German Research Foundation) under Germany's Excellence Strategy – Cluster of Excellence Matter and Light for Quantum Computing (ML4Q) EXC 2004/1 – 390534769 and by the Federal Ministry of Education and Research (BMBF) and the Ministry of Culture and Science of the German State of North Rhine-Westphalia (MKW) under the Excellence Strategy of the Federal Government and the Länder.

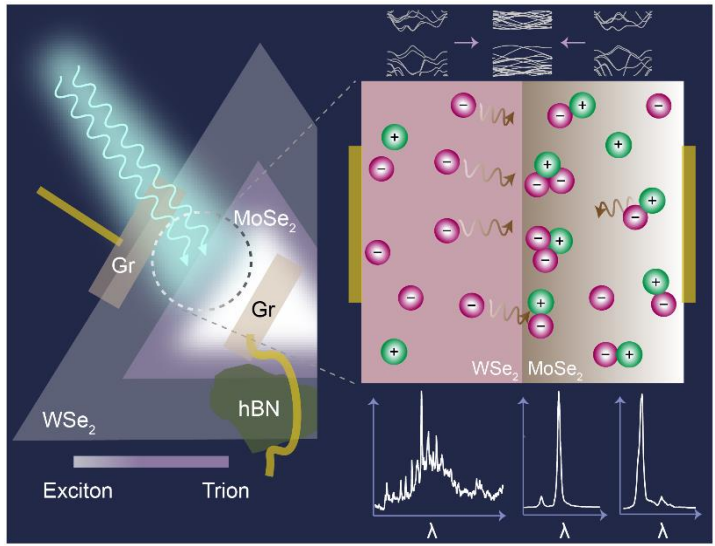

Figure: TOC